\documentclass[%
reprint,
superscriptaddress,
bibnotes,
amsmath,amssymb,
aps,
prb,
]{revtex4-2}

\usepackage[utf8]{inputenc}
\usepackage{graphicx}
\usepackage{dcolumn}
\usepackage{bm}
\usepackage{caption}
\usepackage{subcaption}
\usepackage{multirow}
\usepackage{booktabs}
\usepackage{hyperref}
\usepackage[usenames,dvipsnames]{xcolor}
\usepackage{tikz}
\usepackage{dsfont}
\usepackage{flushend}
\usetikzlibrary{automata, arrows.meta, calc, fit, positioning,shapes}

\usepackage{ragged2e}

\newcommand{\be}{\begin{equation}}
\newcommand{\ee}{\end{equation}}
\newcommand{\bea}{\begin{eqnarray}}
\newcommand{\eea}{\end{eqnarray}}
\newcommand{\rr}{\mathbf{r}}

\newcommand{\m}{\mathrm}
\usepackage{graphicx}
\usepackage{amsmath,amssymb}

\usepackage[normalem]{ulem}

\begin{document}
\author{R. Reho}
 \email{riccardoreho95@gmail.com}
\affiliation{%
 Chemistry Department, Debye Institute for Nanomaterials Science, Condensed Matter and Interfaces, Utrecht University, PO Box 80.000, 3508 TA Utrecht, The Netherland and ETSF
}%
\author{A. R. Botello-M\'endez}
\affiliation{%
 Chemistry Department, Debye Institute for Nanomaterials Science, Condensed Matter and Interfaces, Utrecht University, PO Box 80.000, 3508 TA Utrecht, The Netherland and ETSF
}%

\author{Zeila Zanolli}
\affiliation{%
 Chemistry Department, Debye Institute for Nanomaterials Science, Condensed Matter and Interfaces, Utrecht University, PO Box 80.000, 3508 TA Utrecht, The Netherland and ETSF
}%
             
\title[An \textsf{achemso} demo]{Proximity-Induced Superconductivity in PbTe/Pb
heterostructures from first--principles}


\begin{abstract}
Semiconductor--superconductor interfaces play a crucial role in various applications, including hybrid circuits, thermometry, devices (bolometers, electronic coolers), detectors for high-energy particle physics, and quantum computing as potential hosts of topologically protected zero-energy modes. 
In this work, we  solve the Kohn-Sham Density Functional Theory and Bogoliubov--de Gennes equations to describe the normal and superconducting properties of a PbTe/Pb heterostructure.
We compute the anomalous charge density in real space, estimating its decay length and showing that the pairing potential is anisotropic.
We demonstrate that superconductivity in the PbTe/Pb interface is resilient against strain.
In the normal state we find a large Schottky barrier across the interface, resulting in charge transfer from PbTe to Pb.
We resolve a proximity--induced superconducting gap on the PbTe side, which originates from hybridization between the Pb and PbTe states near the Fermi energy, which occurs in a weak-coupling regime. On the Pb side, the superconducting gap appears partially ‘poisoned’, namely less sharp and less wide than bulk Pb.
Our first-principles simulations provide a quantitative prediction on emergent structural, electronic (charge transfer, potential drops, band offset), and  superconducting properties of the PbTe/Pb interface, namely
key information for the design of devices where the PbTe/Pb interface plays a central role as, for instance, in core/shell PbTe/Pb nanowires.


\end{abstract}

\maketitle
\section{Introduction}\label{sec:intro}
Several types of nanoscale heterostructures (HS) have emerged as promising building blocks for next-generation quantum technologies.
These include planar Josephson junctions ~\cite{li2024selective, yue2024signatures, hell2017two, pientka2017topological}, topological insulator/superconductor interfaces ~\cite{wang2012coexistence, xu2014momentum, xu2014artificial, huang2024emergence}, and core/shell semiconductor/superconductor (SM/SC) nanowires~\cite{lutchyn2010majorana,oreg2010helical}.
%
%
A common feature across these systems is the crucial role of interfacial proximity effects.
%
Strain, band alignment, and external fields can significantly modify the dispersion of states near the Fermi level that participate in superconducting coupling, thus influencing the superconducting properties.
For example, to host Majorana zero modes (MZMs), the interface between SM and  $s$--wave SC
must exhibit 
~\cite{cao2022numerical,paya2024phenomenology}: 
(i)  strong spin-orbit coupling (SOC) and significant Zeeman energy gap ($V_Z$) in the SM, 
(ii) proximity induced superconductivity, and 
(iii) the heterostructure chemical potential must lie within the SM's Zeeman gap and the SC gap.
When the applied Zeeman field exceeds the critical value $V_{C} = \sqrt{\Delta^2+\mu^2}$ the system transitions into a topological phase, hosting MZMs at its ends~\cite{PradaAndreevMajorana2020, aguado2017majorana, mourik2012signatures}. 
Experiments confirmed  
proximity-induced superconductivity in topological insulators (TIs), both in core/shell nanowires~\cite{wang2024gate} and planar junctions~\cite{dong2024proximity}, with the proximity effect decaying exponentially with the TI thickness. 
%
%
PbTe/Pb interfaces are advantageous compared to InAs/Al and InSb/Al ones due to PbTe's narrow bandgap and large Land\'e g-factor, combined with Pb's relatively wide superconducting gap ($\Delta \sim$ 3meV) and critical temperature ($T_C \sim$ 7K)~\cite{ten2022small, jiang2022selective}.

Theoretical models provide insights into the  superconducting properties of both nanowires~\cite{lutchyn2010majorana, oreg2010helical} and planar junctions~\cite{hell2017two, pientka2017topological}.
Usually, model results are presented as a phase diagram of  (topological) superconductivity in terms of several parameters.
However, \textit{first--principles} simulations of Bi$_2$Te$_3$/Nb~\cite{russmann2022proximity} and Bi$_2$Se$_3$/PdTe~\cite{park2020proximity}  emphasize the need of fully including the atomistic details of the interface to accurately compute proximity induced superconductivity.

In this paper, we investigate from \textit{first--principles} proximity effects in a PbTe/Pb HS.
We predict how the structural, electronic (charge transfer, potential drops, band offset), and superconducting properties of the individual materials are modified when they constitute an interface. In particular, we find that the properties of the materials are resilient to strain and doping. All of our findings are obtained using the recently implemented \textsc{SIESTA}-BdG method~\cite{reho2024density, garcia2020siesta}, which simultaneously solves the Kohn-Sham Density Functional Theory and Bogoliubov--de Gennes equations to describe the normal and superconducting properties on the same footing. 
We find significant band bending at the interface,  leading to charge transfer from PbTe to Pb, and affecting the position of the chemical potential. 
In the superconducting state we observe a BCS--like SC gap with a coupling between PbTe and Pb of intermediate strength, namely the states of the HS around the Fermi energy are predominantly of semiconductor character.
We predict a proximity--induced superconducting gap in PbTe and a poisoning of Pb SC gap (with respect to bulk).  
The relevance of such induced superconductivity extends to physical mechanisms where states near the Fermi surface play a central role.
Moreover, we show that the superconducting anomalous charge density $\chi(\rr)$ emerging at the interface is anisotropic in real space.
In particular, the induced SC gap at the interface is resilient under lattice strain. 

\begin{figure*}[t]    
\includegraphics{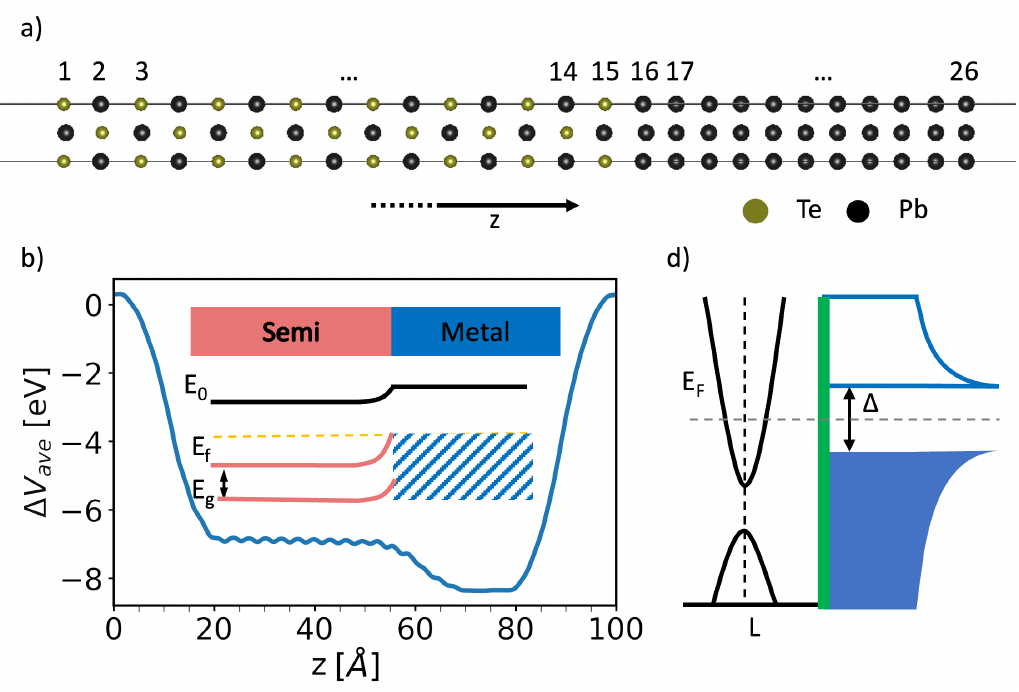}
\caption{(a) Side view of the relaxed (4, -5) PbTe/Pb interface stacked along the [001] direction. 
(b) Average electrostatic potential $\Delta V_{ave}$ across the interface, including a schematic of the energy alignment between a semiconductor and a normal metal. 
(c) Schematic band alignment at the interface between a semiconductor and a superconductor.
} 
\label{fig:pbtepbmodel}
\end{figure*}

\section{PbTe/Pb Heterostructure: Normal state}\label{sec:pbtepb}
PbTe and Pb \{001\} planes are not commensurate. 
Hence, in order to build the HS, it is necessary to introduce strain into the unit cell of both materials.
We employ 15 PbTe and 11 Pb layers, ensuring that the DOS in the central region of PbTe (layers 6--7 Fig~\ref{fig:pbtepbmodel}.(a)] and Pb [layers 21--22 Fig~\ref{fig:pbtepbmodel}.(a)] matches the corresponding bulk values.
We examined a range of strain from $\sim$1\% to $\sim$9\% on the PbTe side, and $\sim$ -8\% to $\sim$ -0.4\% on the Pb side, relaxing the atomic positions for each configuration. 
We denote  strain in the HS by a pair of numbers referring to strain on PbTe and Pb, respectively.
In the following, we focus on the HS with 4\% strain on PbTe and -5\% strain on Pb, namely (4, -5). 

\begin{figure}[t]    
\includegraphics[width=\columnwidth]{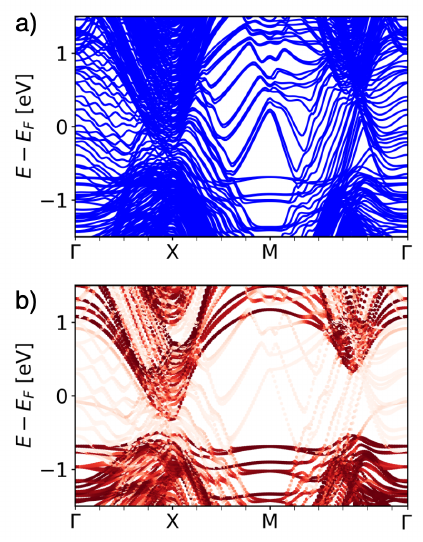}
\caption{
(a) PbTe/Pb normal state band structure.
The bands cross the Fermi level, 
resulting in a metallic heterostructure. 
(b) Normal state bands with the coupling between the PbTe and Pb sides of the heterostructure set to zero. The color intensity represents the contribution from the PbTe orbitals.
\label{fig:pbtepb_bands} 
}
\end{figure}

In the normal state, the bands [Fig.~\ref{fig:pbtepb_bands}.(a)] cross the Fermi level, resulting in a metallic heterostructure  with visible band splitting (mostly near the $M$ point) due to hybridization between PbTe and Pb states, coupling the SC and SM.
To prove this, we remove the interaction between PbTe and Pb by setting their coupling 
to zero in the Kohn Sham Hamiltonian.
The resulting band structure [Fig.~\ref{fig:pbtepb_bands}.(b)]
does not exhibit any splitting. 
We highlight the PbTe contribution to the electronic structure with dark red points in [Fig.~\ref{fig:pbtepb_bands}.(b)], proving that the states near the Fermi energy predominantly belong to the semiconductor.

The interaction between the two materials leads to variation in chemical bonds and relative atomic positions near the interface, along with charge transfer from PbTe to Pb.
Using Mulliken charge analysis~\cite{mulliken1955electronic}, we compute the variation of charge $\Delta Q$ on PbTe ($Q_{PbTe}$) and Pb ($Q_{Pb}$) in the HS 
with respect to their freestanding nominal values.
We find a depletion of charge on PbTe
indicating an electron transfer from the PbTe to the Pb side of the HS (Table~\ref{tab:mulliken} and Fig.\ref{fig:mulliken}). 
The presence of PbTe states crossing the Fermi level and the moderate charge transfer towards Pb, indicate an intermediate-strength coupling regime between the two materials.

When a semiconductor and a metal form an interface, their energy band edges align, creating a potential energy barrier (Schottky barrier) due to the difference in their work functions [Fig.~\ref{fig:pbtepbmodel}(b)].
We calculate from first-principles the average electrostatic potential $\Delta V_{ave}(z)$ ~\cite{baldereschi1988band}, namely the spatial average of all electrostatic interactions in the system. 
The average is performed layer-by-layer, and it quantifies the band alignment in heterostructures.
We predict a significant difference in the average electrostatic potential across the interface $\Delta V_{ave} \sim 1.2$  eV, with the PbTe side being higher in energy than the Pb side [Fig.~\ref{fig:pbtepbmodel}(b)].

\begin{figure}[t]
    \centering
    \includegraphics[width=\columnwidth]{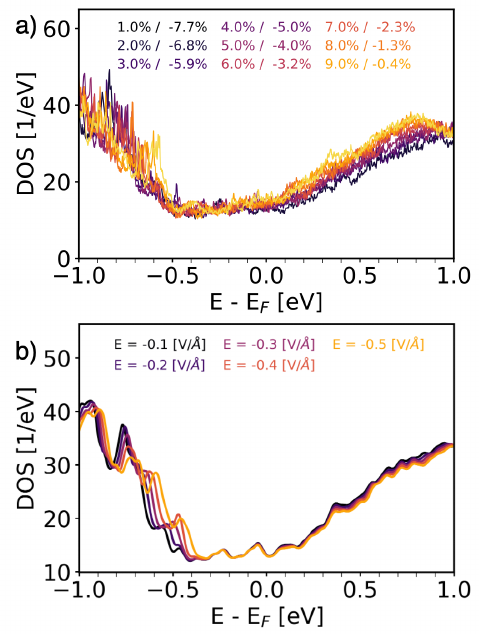}
    \caption{
    PbTe/Pb normal state DOS for different (a) strain and (b) applied electric field across the interface, represented with different colors. 
    }
    \label{fig:normaldosstrain}
\end{figure}

We studied the robustness of these findings against strain and the presence of an external electric field.
Our results indicate that $\Delta V_{ave}$ and the normal state electronic DOS near the Fermi level are not significantly affected by varying strain [Table~\ref{tab:Vavestrain} and Fig.~\ref{fig:normaldosstrain}.(a)].
We compute the  excess charge $\Delta Q$ and observe that it
slightly increases by increasing strain on PbTe (Table~\ref{tab:mulliken}), substantiating the conclusion that the electrostatic environment close to the interface
does not change under lattice deformations. 
Furthermore, the shape and magnitude of the DOS is not affected by the electric field [Fig.~\ref{fig:normaldosstrain}(b)].
The magnitude of the applied electric fields are all considerably above the electrical breakdown point of bulk PbTe which can be estimated to be $6.4\times10^{-6}$ V/\AA~\cite{wang2006relationship}, while its sign and direction are chosen to offset the difference in electrostatic potential at the interface $\Delta V_{ave}$.

\begin{table}[h]
    \caption{ 
    Difference in average electrostatic potential $\Delta V_{ave}$ across the interface of the PbTe/Pb HS for different strain. The (4, -5) HS is highlighted in bold.}
    \begin{tabular}{c|c}
    \hline
     strain (PbTe/Pb) [\%] & $\Delta V_{ave}$ [eV] \\
     \hline
     1.0\%/-7.7\% & 1.36  \\
     2.0\%/-6.8\% & 0.84  \\
     3.0\%/-5.9\% & 0.93  \\
     \textbf{4.0\%}\textbf{/-5.0\%} & \textbf{1.20}  \\
     5.0\%/-4.0\% & 1.08  \\
     6.0\%/-3.2\% & 1.10  \\
     7.0\%/-2.3\% & 1.17  \\
     8.0\%/-1.3\% & 1.22  \\
     9.0\%/-0.4\% & 1.30  \\     
     \hline
    \end{tabular}
    \label{tab:Vavestrain}
\end{table}

\section{PbTe/Pb heterostructure: Superconducting state}

\begin{figure*}[ht]
    \includegraphics[width=0.9\textwidth]{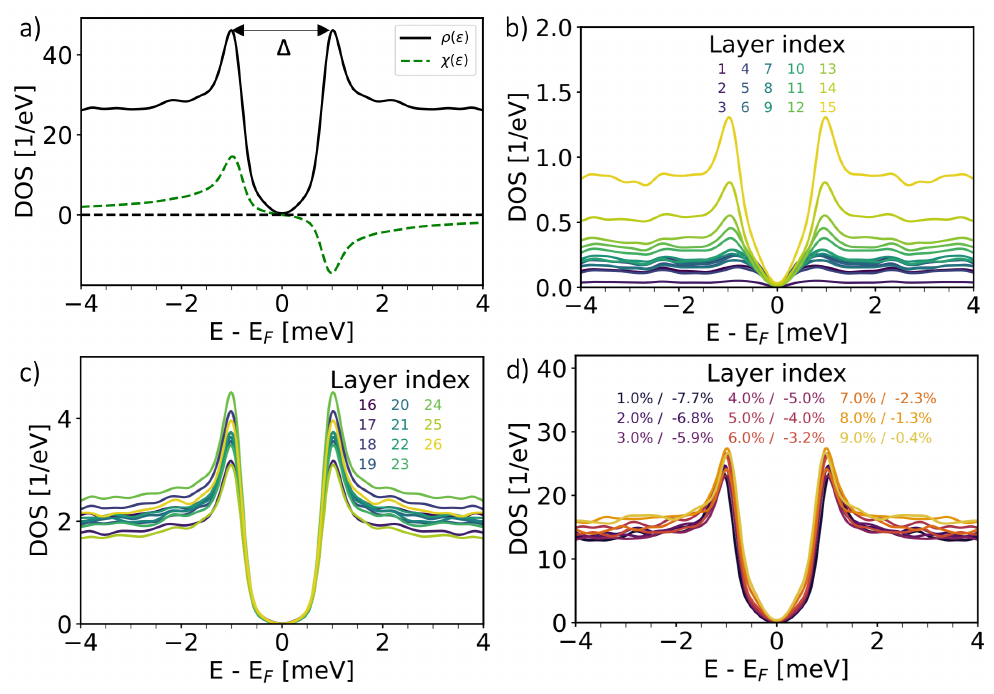}
    \caption{(a) Superconducting DOS (SC--DOS, $\rho(\varepsilon)$, black continuous line) and anomalous DOS (ADOS, $\chi(\varepsilon)$, dashed green line) for the (4, -5) PbTe/Pb HS, with coherence peaks at $\pm \Delta/2 \sim \pm 1$ eV.
    The dashed black line indicates zero DOS.
    SC--DOS projected on individual (b) PbTe and (c) Pb layers, 
    and (d) for the PbTe/Pb HS for diffrent strain values.
    }
    \label{fig:scdos}
\end{figure*}

We used the SIESTA-BdG method to compute the superconducting DOS and band structure 
of PbTe/Pb heterostructures
finding a proximity-induced SC gap, an intermediate-strength coupling between the two materials, and an anisotropic SC gap in real and reciprocal space.
Computational details and geometrical aspects of the HS are presented in Section~\ref{app:compdetails}.

More than fifty years ago, experimental measures~\cite{BlackfordTunnelingInvestigation1969, LykkenMeasurementSuperconducting1971} of bulk Pb revealed an anisotropic U--shaped SC-DOS with two superconducting gaps of width $\sim 2.34-2.5$ and $\sim 2.7-2.8$~meV depending on the crystal orientation. Each coherence peak is split in two, with an energy separation of $\sim0.1-0.2$ meV.
Splitting of the coherence peaks in bulk Pb has been theoretically predicted with SIESTA--BdG~\cite{reho2024density} and KKR-BdG methods~\cite{saunderson2020gap}, and is further discussed in Appendix~\ref{app:bulkpb}.
%
Anisotropies in the superconducting gap are also visible in the BdG spectrum of the HS (Figure~\ref{appfig:bdgspectrum}).  
Indeed, due to heterostructuring, the electrostatic environment of Pb atoms changes layer-by-layer, leading to multiple superconducting gaps at different \textbf{k}--points in the BdG spectrum.

The HS SC gap is U--shaped with coherence peaks $\Delta = 2.0$ meV far apart
[Fig.~\ref{fig:scdos}.(a)], 
reflecting an intermediate-strength coupling regime.
The projected DOS [Fig.~\ref{fig:scdos}.(b) and (c)] show a proximity-induced gap on PbTe and a {\textit{poisoned}} SC gap on Pb, namely a SC gap which 
does not feature a sharp U shape 
and has reduced width with respect to Pb bulk.
The shape of the total DOS is determined by the induced gap in the PbTe side of the heterostucture, as indicated by the projected density of states [Fig.~\ref{fig:scdos}.(b)].
The energy states inside the SC gap are predominantly of PbTe character, hence 
the HS is in the intermediate coupling regime.
The hybridization between Pb and PbTe
``poisons'' the SC gap, leading to a BCS--like superconducting density of states (SC--DOS $\rho(\varepsilon)$).
The Pb side of the heterostructure retains its U-shape but the coherence peaks are less sharp and smaller in magnitude than bulk Pb [Fig.~\ref{fig:scdos}.(c)]. 
In our simulations, the poisoned BCS--like gap at the SM--SC interface is due to structural relaxation of the atomic structure and states hybridization, as we 
do not include magnetic, thermal, or dissipative broadening effects~\cite{takei2013soft}. 
We find that strain only slightly affects the shape of the SC--DOS in the HS [Fig.~\ref{fig:scdos}.(d)].

\begin{figure*}[ht]
    \includegraphics[width=1.0\textwidth]{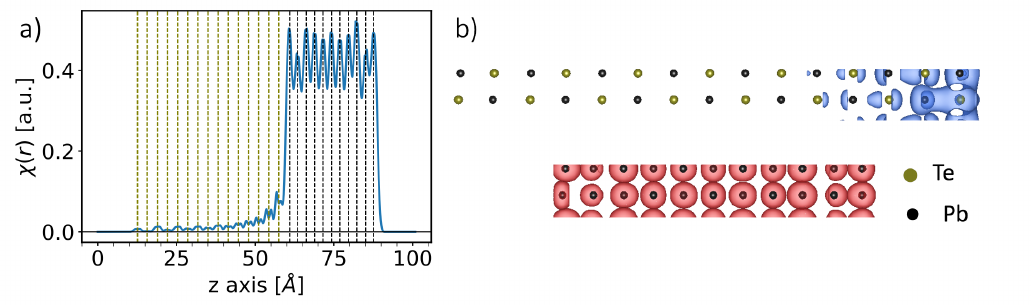}
    \caption{
    (a) Anomalous density of states $\chi(\rr)$ along the z-axis in real space. Vertical dashed lines represent the layer position in the PbTe (olive) and Pb (black) regions of the heterostructure.
    (b) 3D visualization of the anomalous DOS near the interface region. We set the isosurface level to a fixed value ($1.7\times10^{-7}$ 1/Bohr$^3$), which indicate regions where $\chi(\rr)$ is constant. To enhance visualization, the isosurface on the PbTe side of the heterostructure (blue) is set to a value 10 times smaller than the Pb side (red).}    
    \label{fig:chizaxis}
\end{figure*}

The anomalous density of states (ADOS, $\chi(\varepsilon)$)
is dominated by the singlet components (Cooper pair with total spin zero), while the triplet components (Cooper pair with total spin one) are negligible both for buk Pb (Fig.~\ref{fig:bulkpbdosbands}) and the HS (Fig.~\ref{appfig:pbtepbhs_singlettriplet}). We compute the singlet and triplet components following the Balian-Werthamer (BW)~\cite{balian1963superconductivity} representation including SOC. \textit{Saunderson et al.}~\cite{saunderson2025triplet} investigated the difference between solving the BdG equations in the collinear or SOC case. The authors propose an orbital Rashba Edelstein effect that leads to enhanced triplet densities for bulk Pb in the absence of SOC. Notably, a similar weakening of the triplet density in case of SOC is reported here. 

The anomalous charge density $\chi(\rr)$ exhibit atomic scale variations, which we can investigate as a function of the position along the heterostructure.
On the Pb side, $\chi(\rr)$ oscillates around a constant value with maxima at the Pb atomic positions [Fig.~\ref{fig:chizaxis}.(a)]. The oscillations are likely due to the finite size of the simulated system.
On the PbTe side, $\chi(\rr)$ decays from the interface towards the surface. We fitted an exponential function to $\chi(\rr)$ and estimated its decay length to be $\eta\sim14$~\AA.
The estimated decay length $\eta$ is not directly related to the superconducting coherence length of Pb, which is estimated to be $\sim 80$ nm~\cite{poole1999handbook}.
Here, $\eta$ is a result of an effective pairing potential that describes the average build-up of anomalous charge densities in the superconductor.
Finite size effects on superconducting properties of Pb nanofilms~\cite{shanenko2007oscillations,zhang2023proximity} have been explained in terms of quantum well states crossing the Fermi level and the spacing between adjacent layers. In our simulations, the Pb side of the HS is extremely thin compared to the superconducting coherence length of Pb. Consequently, the characteristic energy spacing of quantum well states become significantly larger than the superconducting gap. Quantum confinement might play an important role in the effective electron phonon coupling, and thus the pairing potential. This effect could be included in the initialization of $\Delta$. However, since here we are interested in the interfacial effects, we neglect any quantum size effects.

As discussed above, the superconducting gap $\Delta(\mathbf{k})$ exhibits anisotropy in \textbf{k}-space, meaning its magnitude varies across different \textbf{k}-points.
The anomalous charge density is directly related to the pairing potential by the following expression: 
\begin{equation}
\begin{aligned}
\Delta_{\mu \nu}^{\alpha \beta}(\mathbf{k})=  & -\sum_{l=0,1,2,3} \sum_\gamma i \sigma_l^{\alpha \gamma} \sigma_2^{\gamma \beta} \times \\ &\int_{\mathrm{BZ}} \mathrm{~d} \mathbf{k}^{\prime} \lambda_l^{\mu \nu}\left(\mathbf{k}-\mathbf{k}^{\prime}\right) \chi_{\mu \nu}^l\left(\mathbf{k}^{\prime}\right),
\end{aligned}
\end{equation}
where $\mu$ and $\nu$ label the basis orbitals, $\alpha$ and $\beta$ are spin indices, $\sigma_l$ with $l=0,1,2,3$ are the Pauli matrices, and $\lambda^{\mu\nu}$ are the orbital projected superconducting couplings~\cite{reho2024density}.

We visualize the anomalous charge density in [Fig.~\ref{fig:chizaxis}.(b)], 
where the isosurface level indicate regions with constant $\chi(\mathbf{r})$. 
To enhance visualization, the isosurface on the PbTe side of the heterostructure [blue regions in Fig.~\ref{fig:chizaxis}.(b)] is set to a value 10 times smaller. The anomalous charge density penetrates in the PbTe region, it is enhanced at the interface, and it is highly anisotropic.
On the Pb side of the HS, the anomalous density is homogenous in the central region (layers 21--22) similar to bulk Pb, while it deforms around the interface and the vacuum region. 
Therefore, the pairing potential in the HS is not isotropic.

\section{Computational methods}\label{app:compdetails}
To model the PbTe/Pb interface,
we computed its electronic structure using density functional theory (DFT) as implemented in the \textsc{SIESTA} code~\cite{SolerSIESTAMethod2002,garcia2020siesta}, 
including spin-orbit coupling 
and the PBE exchange-correlation functional~\cite{PerdewGeneralizedGradient1996}.
We employed optimized norm-conserving Vanderbilt pseudopotentials (ONCVPSP~\cite{hamann2013optimized}) in PSML format~\cite{garcia2018psml} from PseudoDojo database~\cite{vanSettenPseudoDojoTraining2018}.
The Kohn-Sham equations were solved using standard double-zeta polarized basis sets of localized atomic orbitals on a $60\times 60\times 1$ Monkhorst-Pack grid. 
Real space integral were evaluated on grids with a cut-off energy of 2000\,Ry for the SC state calculations and 500\, Ry for the normal state ones. 
The Fermi-Dirac occupation function was smoothed using an electronic temperature of 0.1\,meV, to ensure that the broadening of the electronic states is well below the size of the superconducting gap.
All crystal structures were relaxed using the FIRE algorithm with a force threshold of 0.005\,eV/Å 
and a maximum stress tolerance of 0.5\,meV/\AA$^3$.

The superconducting properties of PbTe/Pb were modeled using the \textsc{SIESTA}--BdG method~\cite{reho2024density}. In this approach a semi-empirical superconducting
pairing potential is introduced on top of the self-consistent Kohn-Sham Hamiltonian describing the normal state.  
We employed the \textit{fixed--$\Delta$} solution method, where the BdG Hamiltonian is diagonalized in the presence of a fixed pairing potential $\Delta$. Multiple SCF steps are performed. Self-consistency is achieved on the normal state Hamiltonian, assuming that perturbations induced by the self-consistent update of the anomalous charge are negligible. In the \textit{fixed--$\Delta$} solution method, the pairing potential $\Delta$ is a semi-empirical parameter. The initial value of the pairing potential is expressed, in real space via a superconducting strength parameter $\bar{\Delta}$ as $\Delta(\mathbf{r}) = \bar{\Delta}g(\mathbf{r})$, with $g(\mathbf{r)}$ a general function. We refer to this approach as the \textit{superconducting strength representation}.
This real-space approach is useful to model heterostructures such as the one studied in this work.
In the case of the (4,-5) PbTe/Pb HS we initialize the superconducting pairing potential as touching spherical hardwells with radius $\mathbf{r} =1.58$~\AA~and strength $\bar{\Delta} = 1.50$~meV around Pb atoms. 

Our approach utilizes an effective pairing potential that describes the average build-up of anomalous charge densities in the superconductor, providing a robust framework for interpreting the physical phenomena. Therefore, we do not expect large variations on the superconducting properties of the HS as the Pb thickness changes, even though finite-size effects can significantly influence the SC-DOS in planar Josephson junctions.~\cite{aceves2024superconductivity,yamazaki2024quantum,shanenko2007oscillations}
We demonstrate that a proper initialization of the superconducting pairing potential is essential to avoid unphysical results (Appendix~\ref{app:initpairpot}, Fig.~\ref{fig:pbtepb_realvsintravsfull}). 
Moreover, we demonstrate how the \textsc{SIESTA}-BdG method can be used to investigate various pairing mechanisms and their effects on the superconducting density of states (SC--DOS).

\section{Conclusion}\label{sec:conclusion}
We used a combined Kohn--Sham and Bogoliubov--de Gennes approach to  
investigate the semiconductor--superconductor (SM--SC) heterostructure composed of PbTe and Pb, focusing on the interfacial properties. 
We discussed the (normal state) metallization of the heterostructure driven by strong hybridization between PbTe and Pb states at the interface. 
We predict a significant difference in electrostatic potential at the interface, which poses a challenge in achieving the necessary energy alignment between the SM Zeeman gap and SC gap.
These findings remain robust against strain and electric field.

In the superconducting state, a BCS--like superconducting gap is computed, indicating a intermediate superconducting coupling between the two systems. 
Moreover, we predicted proximity--induced superconductivity in the PbTe region of the HS. We resolved the superconducting density of states layer by layer, revealing that the induced SC gap is enhanced near the interface, with a poisoned SC gap on the Pb side.
We analyzed the shape and decay of the anomalous charge density $\chi(\rr)$, revealing an anisotropic pairing potential. 
We observed a weak superconducting coupling regime at the interface, which prevents the superconductor to overscreen the HS properties. 
The weak regime favours the device's tunability with gate and bias voltages, creating a route to get the desirable energy alignment between the two materials. 

The field of topological quantum materials is increasingly moving towards identifying optimal design parameters to control and enhance the functionality of hybrid devices.
Our results naturally fit in this direction, by providing first-principles quantitative predictions  of the superconducting and electronic properties of the PbTe/Pb semiconductor/superconductor interface and their resiliance to strain and electric field. 
We expect our findings to benefit both experimentalists and theoreticians alike, due to the key role played by this interface in various technologically relevant devices, among which core/shell semiconductor/superconductor nanowires.

\section*{Acknowledgments}\label{sec:acnkowledgements}
The authors acknowledge the fruitful discussion with Erik Bakkers, Arnold H. Kole and Nils Wittemeier.
ZZ acknowledges the research program “Materials for the Quantum Age” (QuMat) for financial support. This program (registration number 024.005.006) is part of the Gravitation program financed by the Dutch Ministry of Education, Culture and Science (OCW).
RR and ZZ acknowledge financial support from Sector Plan Program 2019-2023. 
This work was sponsored by NWO-Domain Science for the use of supercomputer facilities. 
This publication is part of the project ”Quantum Materials by Design” with file number 2024.012 of the research programme Computing Time on National Computer Facilities which is (partly) financed by the Dutch Research Council (NWO) under the grant https://doi.org/10.61686/JUXDK41645 .
We also acknowledge that the results of this research have been achieved using the Tier-0 PRACE
Research Infrastructure resource Discoverer based in Sofia, Bulgaria  (OptoSpin project id. 2020225411). 
This project has received funding from the European Union’s Horizon Europe research and innovation program under Grant Agreement No 101130384 (QUONDENSATE).

\section{Data availability}
The data of this work are available at the following NOMAD repository~\cite{nomad}.

\section{keywords}
proximity interaction, superconductivity, semiconductor-superconductor heterostructures, first-principles simulations, Bogoliubov de Gennes equations

\appendix
\renewcommand{\thesection}{\Alph{section}}
\renewcommand{\thesubsection}{\arabic{subsection}}
\setcounter{figure}{0}
\renewcommand{\figurename}{Figure}
\renewcommand{\thefigure}{\Alph{section}\arabic{figure}}
\setcounter{equation}{0}
\renewcommand{\theequation}{\Alph{equation}}
\renewcommand{\thetable}{\Alph{table}}
\setcounter{table}{0}

\section{Geometrical aspect of the HS}\label{app:geomaspect}
We constructed the PbTe/Pb HS by stacking 15 layers of PbTe and 11 layers of Pb along the [001] direction.
The HS extends infinitely in the x and y directions. Periodic replicae are separated by  26~\AA~of vacuum along the z--axis.
The PbTe slab was constructed by rotating the in-plane conventional cell by 45 degrees and reconstructing its primitive cell.
Farey sequences~\cite{niven1991introduction} were employed to identify commensurate cells for the full heterostructure.
This method inherently introduces strain in the system.
We considered several strained configurations of the PbTe/Pb heterostructure.
The strain on each side of the heterostructure is defined as:
\be
\text{strain}= \frac{\m{a}_{\m{hs}}-\m{a}_{\m{rlx,i}}}{\m{a}_{\m{rlx,i}}} \quad \m{i = Pb,PbTe}
\ee
where $\m{a}_{\m{hs}}$ is the lattice parameter of the heterostructure while $\m{a}_{\m{rlx,i}}$ is the relaxed lattice parameter of the constituent materials.

\section{Bulk Pb and bulk PbTe}\label{app:bulkPbPbTe}
\renewcommand{\thesection}{\Alph{section}}
\renewcommand{\thesubsection}{\arabic{subsection}}
\setcounter{figure}{0}
\renewcommand{\figurename}{Figure}
\renewcommand{\thefigure}{\Alph{section}\arabic{figure}}
\setcounter{equation}{0}
\renewcommand{\theequation}{B}
\renewcommand{\thetable}{\Alph{section}\arabic{table}}
\setcounter{table}{0}
We present the crystal structure and electronic properties of bulk Pb and PbTe, emphasizing the orbitals' contribution to the electronic states near the Fermi level.
The interaction and hybridization between these states are critical for the proximity--induced superconductivity effect at the PbTe/Pb interface.

\subsection{Bulk Pb}\label{app:bulkpb}
\begin{figure}[t]    
    \centering
    \includegraphics[width=0.9\columnwidth]{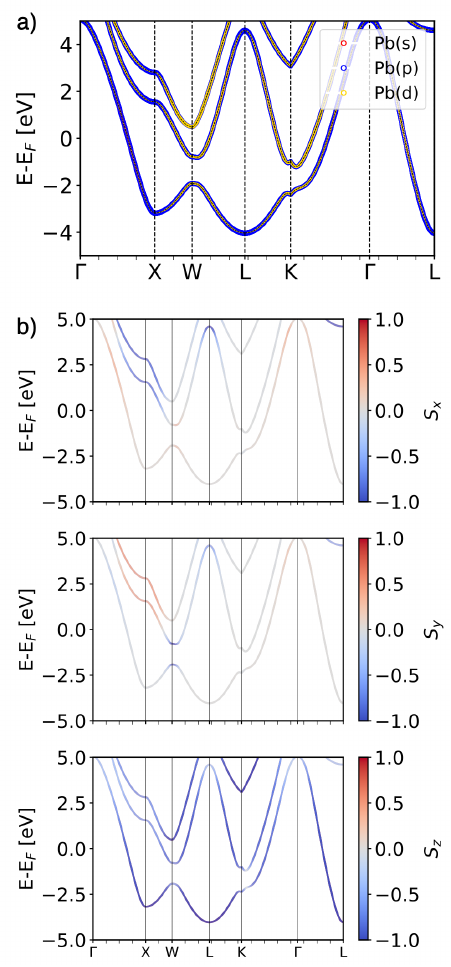}
    \caption{\label{fig:bulkpbspinbands}
    Orbital (a) and spin (b) projected band structure of bulk Pb in the normal state. 
    In (a) we observe that the electronic states crossing the Fermi level are $p$ and $d$ orbitals. 
    In (b) we observe a spin flip at $W$ for  $S_x$ and $S_y$ spin orientations. 
}
\end{figure}

The space group of Pb is Fm$\bar{3}$m. Structural relaxation yielded a primitive lattice vector of $a_{prim}=3.547$\AA~($a_{conv}=5.016$ \AA).
Pb is a metal with an orbital character dominated by $d$ and $p$ orbitals close to the Fermi level at the high-symmetry points $W$ and $K$ [Fig.~\ref{fig:bulkpbspinbands}.(a)]. The computed total spin moment is zero, in agreement with bulk Pb being not magnetic. 
We investigate the spin projection for the electronic states close to the Fermi level, as those are relevant for the formation of Cooper pairs. The spin--projected band structure [Fig.~\ref{fig:bulkpbspinbands}.(b)] show that the spin components $S_{x}$ and $S_{y}$ change sign at the $W$ point (spin flip).  

\begin{figure*}[t]    
\includegraphics[width=1.0\textwidth]{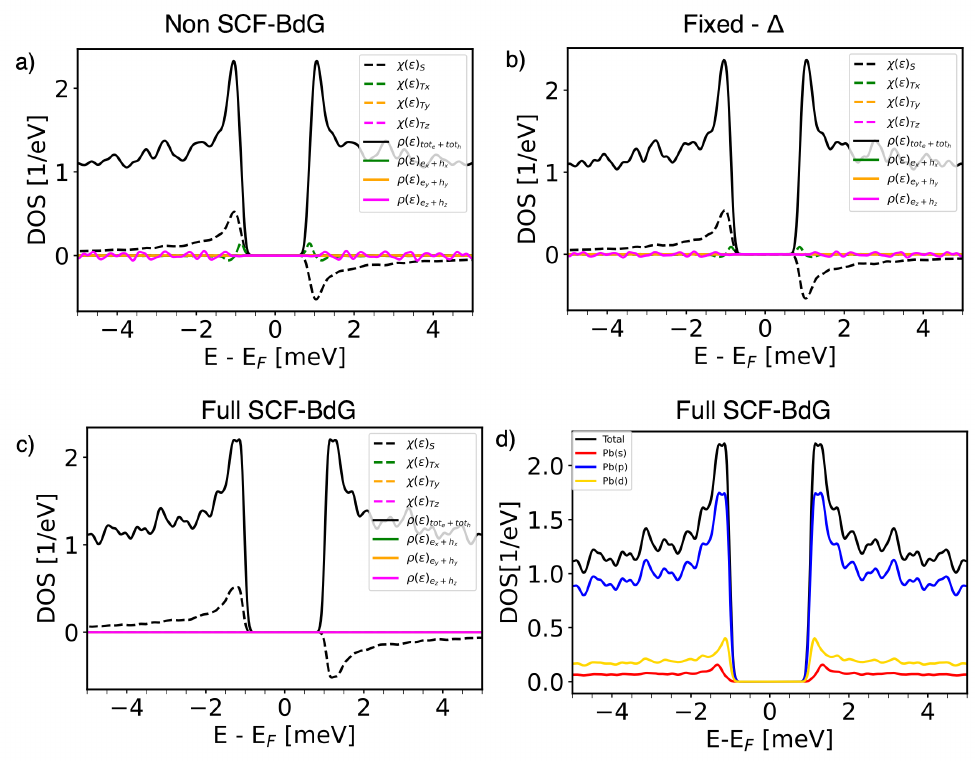}
\caption{
SC--DOS $\rho(\varepsilon)$ (continous lines) and ADOS $\chi(\varepsilon)$ (dashed lines) for bulk Pb computed with the solution methods (a) \textit{non SCF-BdG}, (b) \textit{fixed-$\Delta$}, and (c) \textit{full SCF-BdG}. Panel
(d) is the orbital projected normal density of states. Some lines at zero DOS are not visible as they are hidden behind by the magenta full line.
\label{fig:bulkpbdosbands}
}
\end{figure*}

We computed the superconducting state using the \textsc{SIESTA}-BdG code~\cite{reho2024density}, including SOC and employing three solution methods: \textit{non SCF-BdG}, \textit{fixed-$\Delta$}, and \textit{full SCF-BdG}.
All three solution methods predict a conventional U-shaped SC gap of $\sim$3 meV (Fig.~\ref{fig:bulkpbdosbands}).

In the following, we discuss the components of the normal $\rho$ and anomalous $\chi$ densities, highlighting how those are obtained from the projected densities of states $\rho^{\alpha\beta}_{\mu\nu}$ and $\chi^{\alpha\beta}_{\mu\nu}$ computed by the code.

The projected densities of states are given by~\cite{reho2024density}

\begin{equation} \label{eq:pdos_anpdos}
    \begin{aligned}
        \rho_{\mu \nu}^{\alpha \beta}(\mathbf{k})= & \sum_i f\left(\varepsilon_i(\mathbf{k})\right) u_{i \mu}^{* \alpha}(\mathbf{k}) u_{i \nu}^\beta(\mathbf{k}) \\
        & +\sum_i \bar{f}\left(\varepsilon_i(-\mathbf{k})\right) v_{i \mu}^\alpha(-\mathbf{k}) v_{i \nu}^{* \beta}(-\mathbf{k}), \\
        \chi_{\mu \nu}^{\alpha \beta}(\mathbf{k})= & \sum_i f\left(\varepsilon_i(\mathbf{k})\right) u_{i \nu}^\beta(\mathbf{k}) v_{n \mu}^{* \alpha}(\mathbf{k}) \\
        & +\sum \bar{f}\left(\varepsilon_i(-\mathbf{k})\right) u_{i \mu}^\alpha(-\mathbf{k}) v_{i \nu}^{* \beta}(-\mathbf{k}).
    \end{aligned}
\end{equation}
The tensor structure describes how the weight is distributed among orbitals and spin degrees of freedom. Each projected density of states contains an `electron'' (proportional to $f(\varepsilon_i({\mathbf{k}}))$) and `hole'' (proportional to $\bar{f}(\varepsilon_i({\mathbf{k}}))$) contribution, corresponding to the two components of the BdG quasiparticles.
The spin sector of $\rho^{\alpha\beta}_{\mu\nu}$ is expanded in the basis of Pauli matrices ($\sigma^0, \sigma^x, \sigma^y, \sigma^z $). This yields one spin-summed term and a three components term, conventionally labelled as $x,y,z$.
For the anomalous density $\chi^{\alpha\beta}_{\mu\nu}$ the same decomposition is rearranged into singlet (S) and triplet (T) components according to the Balian-Werthamer representation~\cite{balian1963superconductivity}.
Summing over all orbital and spin indices one recovers the scalar density of states ($\rho (\varepsilon)$ and $\chi(\varepsilon)$).

In Fig.~\ref{fig:bulkpbdosbands} we report the total (spin-summed) normal density as sum of the electron and hole components  
$\rho_{tot_e}+\rho_{tot_h}$,
its $i = x, y, z$ spin-components 
$\rho_i = \rho_{e_i} + \rho_{h_i}$, 
and the singlet and triplet components of the anomalous density $\chi_{S}$ and $\chi_{T}$.

The z component of the SC density of states $\rho(\varepsilon)_z$ exhibits small oscillations around zero when computed using the \textit{non-SCF-BdG} and \textit{fixed-$\Delta$} methods but vanishes when computed using the \textit{full SCF-BdG} method, which is the most accurate one [Fig.~\ref{fig:bulkpbdosbands}.(a-c)]. The x and y components of the normal density are always identically zero, regardless of the solution method employed. 
The triplet components 
$\chi(\varepsilon)_{T_x}$ is almost zero for \textit{non SCF-BdG} and \textit{fixed-$\Delta$} solution methods, and identically zero for the \textit{full SCF-BdG one}.
The $\chi(\varepsilon)_{T_y}$ and $\chi(\varepsilon)_{T_z}$ components of $\chi$ are always zero. 
The small oscillations around zero for $\rho(\varepsilon)$ and $\chi(\varepsilon)$ are due to numerical noise. The oscillations appear only for the \textit{non SCF-BdG} and \textit{fixed-$\Delta$} solution methods because in these cases the anomalous components are not updated, namely they are kept constant with respect to the initial assumption. 
This further validates our implementation and explains why the DOS of triplet components of $\chi(\varepsilon)$ are symmetrical with respect to the $E_F$ in the \textit{non SCF-BdG} and \textit{fixed- $\Delta$} cases.

We initialize the pairing potential in the \textit{superconducting strength representation}
(on a real-space grid) as touching spherical hardwells with radius $\mathbf{r} = 1.58$~\AA.
In the \textit{non SCF--BdG} and \textit{fixed--$\Delta$} methods, the initial strength of the superconducting pairing potential is $\bar{\Delta} = 1.50$~meV. In the \textit{full SCF-BdG} method, instead, the initial value of the superconducting coupling is $\lambda(\mathbf{r}) = 80.3$~eV.
We computed the orbital projected superconducting density of states  [Fig.~\ref{fig:bulkpbdosbands}.(d)] and conclude that Pb $p$ and $d$ are the dominant electronic states close to the Fermi level, and  hence relevant for the superconducting properties. 
The \textit{full SCF-BdG} predicts that the coherence peak at $\sim \pm 1.8$~meV displays a small splitting. 
This splitting is physical and has been experimentally measured in bulk Pb~\cite{BlackfordTunnelingInvestigation1969, LykkenMeasurementSuperconducting1971}, and found to be $\sim 0.1-0.2$~meV depending on the crystal direction. 
Simulations with the KKR-BdG method have analyzed the gap anisotropy and identified three superconducting gaps~\cite{saunderson2020gap}.
In our earlier work~\cite{reho2024density}, we  report two superconducting gaps in bulk Pb, consistently with these findings. Accurately resolving the splitting of the coherence peak requires an extremely low smearing temperature ($T = 0.002$ meV).
We note that the coherence peak splitting is only predicted within the \textit{full SCF-BdG} method:
the splitting is not present when the SC-DOS is computed with the \textit{fixed-$\Delta$} and \textit{non scf-BdG} methods.

\subsection{Bulk PbTe}\label{app:bulkpbte}
\begin{figure}[t]
    \centering
    \includegraphics[width=1.0\columnwidth]{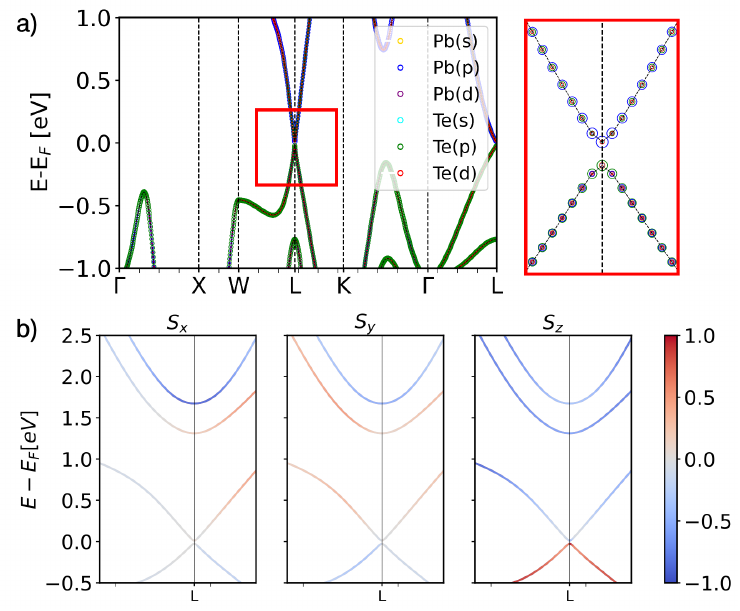}
    \caption{
    Electronic band structures for PbTe bulk.
    (a) Orbital projected band structure. The inset shows an energy region close to the Fermi level.
    (b) Spin projected band structure. Positive and negative spins are depicted in red and blue, respectively.
    }
    \label{fig:bulkpbtebands}
\end{figure}

The space group of PbTe is Fm$\bar{3}$m.
Structural relaxation yielded a primitive lattice vector of $a_{prim}=4.578$ \AA~(to be compared with the experimental value $4.568$ \AA~\cite{lide2004crc,lawson1951method}).
PbTe is a semiconductor. Our calculations predict its spin-orbit coupling gap to be $0.023$ eV in the ground state~\cite{aguado2020gw}. The magnitude of the gap varies with strain and can reach hundreds of meV~\cite{aguado2020gw} as showed in Table~\ref{tab:pbtegapstrain}.
The states close to the Fermi level are predominantly composed of Te $p$--orbitals in the valence manifold (with minor contributions from Pb $s,p,d$--orbitals) and Pb $p$--orbitals in the conduction manifold [Figs.~\ref{fig:bulkpbtebands}.(a,b)].
Near the Fermi level at $L$, the $S_x$ $S_y$ spin components of the bands change sign while we observe a spin flip along the $S_z$ component.
For energies $1.5$ eV above the Fermi level, we observe a $\sim0.5$ eV SOC splitting in the $S_x$ and $S_y$ components [Fig.~\ref{fig:bulkpbtebands}.(b)].

\begin{table}[b]
    \caption{\label{tab:pbtegapstrain} 
    Electronic gap ($\Delta E_{L}$) of PbTe as a function of its in-plane lattice parameter. The configuration that matches the value of the lattice parameter for the (4,-5) PbTe/Pb HS is highlighted in bold.}
    \begin{tabular}{c|c}
    \hline
     Lattice parameter [\AA] & $\Delta E_{L}$ [eV] \\
     \hline
     4.39 & 0.251  \\
     4.44 & 0.224  \\
     4.49 & 0.147  \\
     4.53 & 0.069 \\
     \textbf{4.58} & \textbf{0.023}  \\
     4.63 & 0.058  \\
     4.67 & 0.107  \\
     4.72 & 0.149 \\
     4.76 & 0.181  \\
     \hline
    \end{tabular}
\end{table}

\begin{figure}[t]
    \centering
    \includegraphics[width=1.0\columnwidth]{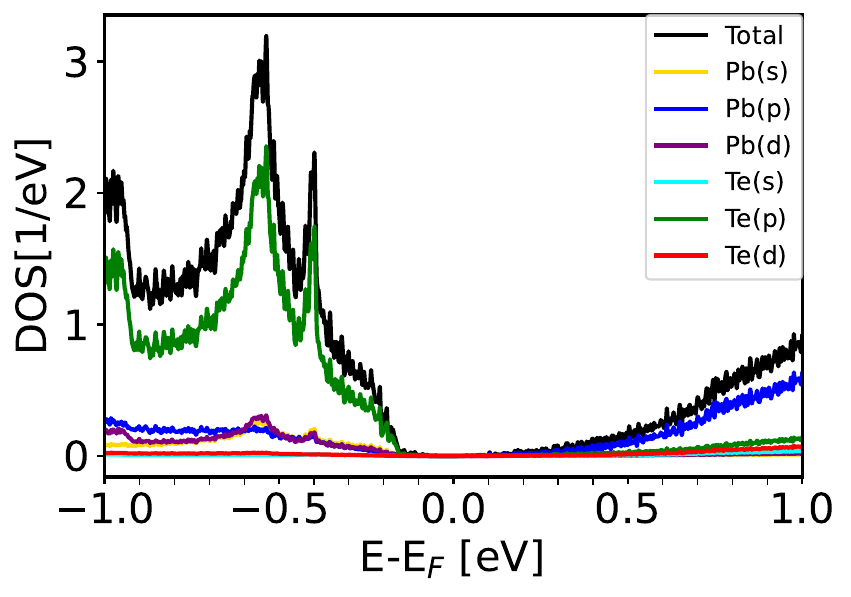}
    \caption{
    Projected DOS of bulk PbTe.
    Near the Fermi level, the valence band is predominantly influenced by Te $p$ orbitals, whereas the conduction band shows a substantial contribution from Pb $p$ orbitals
    }
    \label{fig:fatdospbte}
\end{figure}


\section{PbTe/Pb HS: Initialization of the Pairing Potential and Its Impact on the Superconducting Properties}\label{app:initpairpot}

\begin{figure}[b]
    \centering
    \includegraphics[width=1.0\columnwidth]{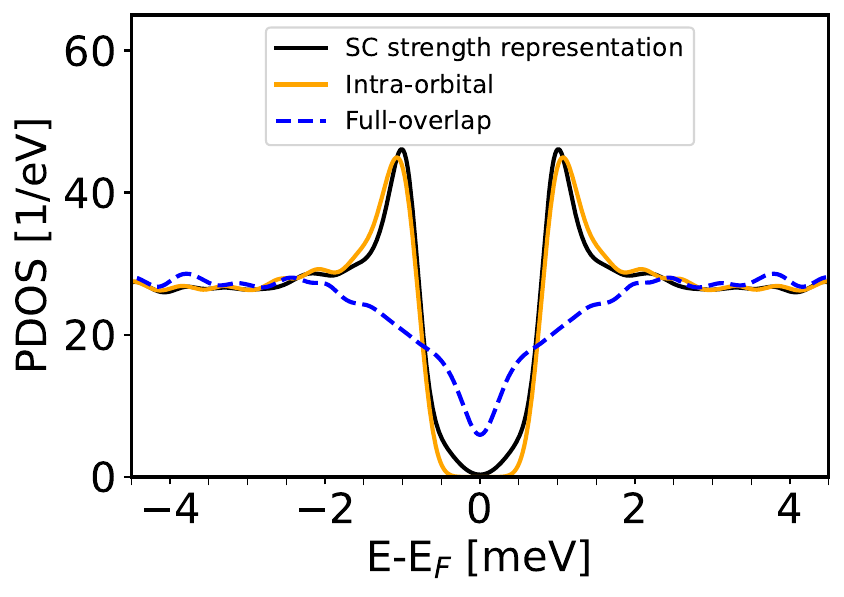}
    \caption{\label{fig:pbtepb_realvsintravsfull}
    The SC--DOS, $\rho(\varepsilon))$, of the (4,-5) PbTe/Pb HS is computed using the \textit{superconducting strength representation method} (solid black line) and the \textit{orbital representation} method. In the orbital representation, we evaluate intra-orbital coupling (solid orange line) and full-overlap coupling (dashed blue line). When pairing occurs between electrons and holes within the same orbital on the same atom (intra--orbital), a conventional U-shaped gap emerges. In contrast, when superconducting coupling involves interactions between different orbitals across different atoms (superconducting strength representation or full--overlap), the hard superconducting gap becomes poisoned.
    }
\end{figure}

In the \textsc{SIESTA}-BdG method one can employ different methods for solving 
the BdG equations and different approaches 
for initializing the pairing potential $\Delta$~\cite{reho2024density}. 
We tested multiple initializations of the pairing potential.
In the main text, we report results obtained within the superconducting
strength representation for all simulations. 

We compared the superconducting strength and orbital representation
(\textit{intra--orbital} coupling or \textit{full--overlap}) for the PbTe/Pb heterostructure (Fig.~\ref{fig:pbtepb_realvsintravsfull}). The intra--orbital coupling lacks the interactions between different orbitals that can be captured by the superconducting strength representation, making it more localized. As a result, the \textit{intra--orbital} coupling misses the hybridization and coupling between states belonging to the PbTe and Pb sides of the heterostructure, leading to a U--shaped DOS. However, when we excessively couple states belonging to different orbitals and atoms (full--overlap), the superconducting gap closes. Therefore, initializing the pairing potential in real--space as touching sphere hardwells is a physically intuitive and reasonable approach.

\section{Additional figures and tables}\label{app:mulliken}
\setcounter{table}{0}

\begin{figure}[h]
    \includegraphics[width=1.0\columnwidth]{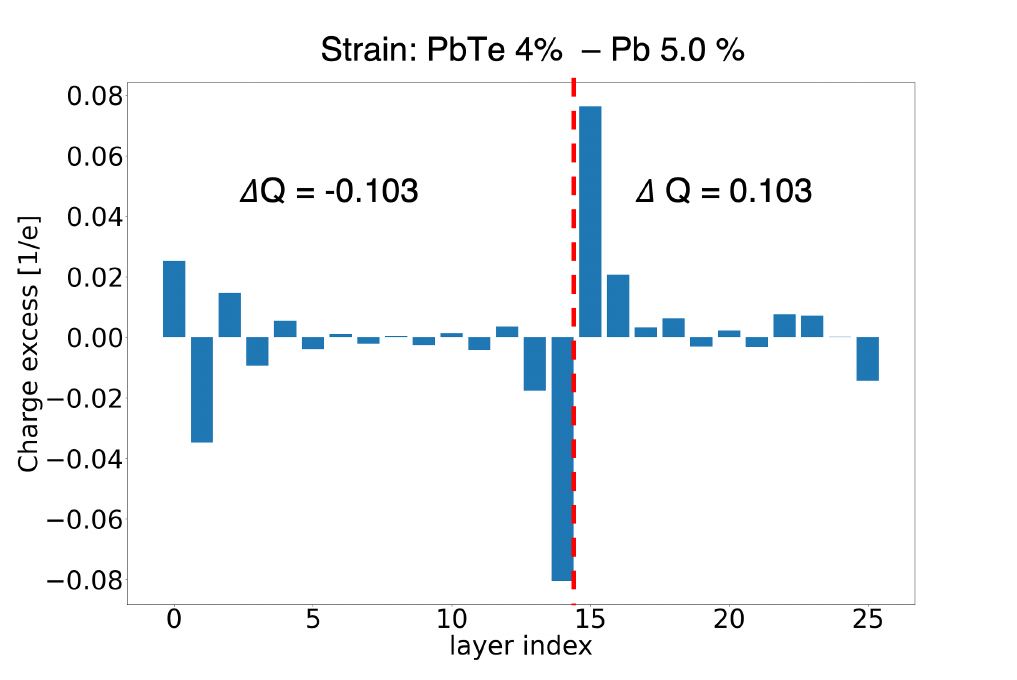}
    \caption{\label{fig:mulliken}
    Mulliken charge analysis for (4,-5) PbTe/Pb HS. Layer-by-layer excess charge. Positive (negative) values of $\Delta Q$ denote the acceptor (donor) character of the layer. The red dashed line denote the separation between the PbTe and Pb side of the heterostructure.
    The sum of the excess charge on the PbTe and Pb side $\Delta Q$ is reported in the plot.
    }
\end{figure}

\begin{figure}[t]
    \centering
    \includegraphics[width=1.0\columnwidth]{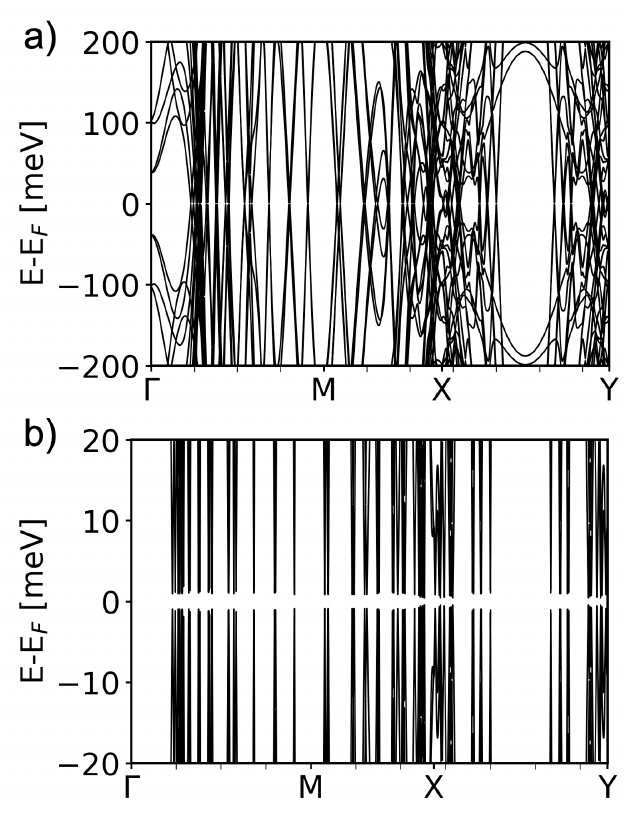}
    \caption{BdG spectrum for the (4,-5) PbTe/Pb HS. The right panel displays the superconducting states dispersion in a narrower energy range.}
    \label{appfig:bdgspectrum}
\end{figure}

This sections presents additional figures and tables related to the PbTe/Pb HS.
We performed a Mulliken charge analysis and calculated the excess charge $\Delta Q$ under different strain conditions, finding that $\Delta Q$ remains robust agains strain variations. (Table~\ref{tab:mulliken}).
A layer-by-layer decomposition of $\Delta Q$ (Fig.~\ref{fig:mulliken}) reveals charge depletion on the PbTe side of the interface, with a corresponding accumulation on the Pb side.
Fig.~\ref{appfig:bdgspectrum} shows the BdG spectrum of the (4,-5) PbTe/Pb HS. The SC gap $\Delta(\mathbf{k})$ is anisotropic, as evidenced by the variations in the width of the splitting of the BdG dispersion along the Brillouin Zone

\begin{figure}[ht]    
\includegraphics[width=1.0\columnwidth]{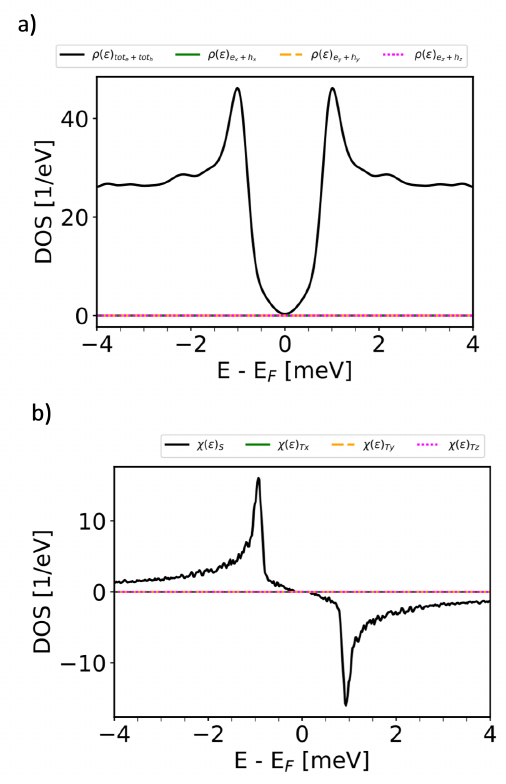}
\caption{(a) SC--DOS $\rho(\varepsilon)$ and (b) ADOS $\chi(\varepsilon)$ for the (4,-5) PbTe/Pb HS computed with the \textit{fixed-\textit{$\Delta$}} solution method. The triplet components of $\chi(\varepsilon)$ are clearly negligible.
\label{appfig:pbtepbhs_singlettriplet}
}
\end{figure}

For the PbTe/Pb (4,-5) HS, we compared the components of $\rho(\varepsilon)$ and $\chi(\varepsilon)$ following the approach discussed for bulk Pb in Appendix~\ref{app:bulkpb}. Using the \textit{fixed-$\Delta$} solution method, we observed a similar behaviour as for bulk Pb and concluded that the triplet components of the anomalous density can also be neglected in the HS (Fig.~\ref{appfig:pbtepbhs_singlettriplet}).

\begin{table}[ht]
    \caption{\label{tab:mulliken} 
    Excess of charge $\Delta Q$ computed for different strain. 
    Positive (negative) values of $\Delta Q$ denote the acceptor (donor) character of the layer.
    We compare the excess of charge on the PbTe side ($\Delta Q_{PbTe}$) of the heterostructure, with the one on the Pb side ($\Delta Q_{Pb}$). The (4, -5) HS is highlighted in bold.}
    \begin{tabular}{c|c|c}
    \hline
     strain (PbTe/Pb) [\%] & $\Delta Q_{PbTe}$ [1/e] & $\Delta Q_{Pb}$ [1/e] \\
     \hline
     1.0\%/-7.7\% & -0.096 & 0.096  \\
     2.0\%/-6.8\% & -0.098 & 0.098 \\
     3.0\%/-5.9\% & 0.101 & 0.101  \\
     \textbf{4.0\%}/\textbf{-5.0\%} & \textbf{-0.103} & \textbf{0.103}  \\
     5.0\%/-4.0\% & -0.107 & 0.107  \\
     6.0\%/-3.2\% & -0.106 & 0.106  \\
     7.0\%/-2.3\% & -0.109 & 0.109  \\
     8.0\%/-1.3\% & -0.114 & 0.114  \\
     9.0\%/-0.4\% & -0.119 & 0.119  \\     
     \hline
    \end{tabular}
\end{table}


\clearpage
\thispagestyle{empty}
\bibliography{bibliography}

\end{document}